\documentclass[acmsmall,screen]{acmart}
\usepackage{amsmath} % acmart already supplies the math fonts and symbols.
\usepackage{algorithmic}
\usepackage{graphicx}
\usepackage{textcomp}
\usepackage{xcolor}
\usepackage{booktabs}
\usepackage[skins]{tcolorbox}
\usepackage{multirow}
\usepackage{listings}
\usepackage{colortbl}
\usepackage{array}
\usepackage{tabularx}
\usepackage{pifont}
\usepackage{pgfplots}
\usepackage{threeparttable}
\usepackage{placeins}
\usepackage{xspace}
\usepackage{enumitem}
\usepackage{makecell}
\newcommand{\papertablestyle}{%
  \footnotesize
  \setlength{\tabcolsep}{4pt}%
  \renewcommand{\arraystretch}{1.12}%
}

\begin{document}

\title{Reasoning Patterns for Efficient AI Coding: An Empirical Study
}
%Reasoning Patterns Matter: An Empirical Study of LLM Coding for Inference Efficiency
%Reasoning Patterns in AI Coding: An Empirical Study and Pattern-Aware Inference Efficiency

\author{Zhe Yin}
\affiliation{%
  \institution{Shanghai Jiao Tong University}
  \city{Shanghai}
  \country{China}
}
\email{yin\_zhe@sjtu.edu.cn}

\author{Yuling Shi}
\affiliation{%
  \institution{Shanghai Jiao Tong University}
  \city{Shanghai}
  \country{China}
}
\email{yuling.shi@sjtu.edu.cn}

\author{Xiaodong Gu}
\affiliation{%
  \institution{Shanghai Jiao Tong University}
  \city{Shanghai}
  \country{China}
}
\email{xiaodong.gu@sjtu.edu.cn}

\author{Beijun Shen}
\authornote{Corresponding author.}
\affiliation{%
  \institution{Shanghai Jiao Tong University}
  \city{Shanghai}
  \country{China}
}
\email{bjshen@sjtu.edu.cn}

\begin{abstract}
Large Reasoning Models (LRMs) generate long reasoning traces that expose intermediate inference behavior, offering opportunities to study and control reasoning in AI coding. However, the underlying patterns in these traces and their potential for analyzing and guiding coding models remain poorly understood, limiting systematic characterization and inference-time control of LRM reasoning.
We conduct an empirical study of reasoning patterns across four representative coding tasks. We transform reasoning traces into structured action sequences and mine recurring fragments using within-task support, cross-task contrasts, and valid--invalid support contrasts. We characterize four general reasoning patterns by coding agents: design-to-implementation for code generation, state tracing for execution reasoning, evidence--fix--recheck for debugging, and semantic mapping for translation. We further identify positive and negative patterns associated with valid and invalid reasoning traces, respectively. Positive patterns reflect verification-driven convergence, whereas negative patterns reflect repeated design revision, reflection, and knowledge recall without closure.
To demonstrate their practical utility, we evaluate three pattern-based applications: task identification, pattern-guided prompting, and pattern-aware early stopping. Features capturing general reasoning patterns enable accurate task identification from reasoning traces without accessing task descriptions or final answers. Prompting guided by positive and negative patterns improves the overall accuracy of downstream tasks, while negative pattern-aware early stopping substantially reduces API-reported token usage with only a modest decrease in pass@1. All three applications also generalize to previously unseen models, showing that reasoning patterns can support both the analysis and control of LRM behavior in AI coding.
\end{abstract}
\begin{CCSXML}
<ccs2012>
   <concept>
       <concept_id>10010147.10010178</concept_id>
       <concept_desc>Computing methodologies~Artificial intelligence</concept_desc>
       <concept_significance>300</concept_significance>
       </concept>
   <concept>
       <concept_id>10011007.10011074.10011092</concept_id>
       <concept_desc>Software and its engineering~Software development techniques</concept_desc>
       <concept_significance>300</concept_significance>
       </concept>       
 </ccs2012>
\end{CCSXML}

\ccsdesc[300]{Software and its engineering~Software development techniques}
\ccsdesc[300]{Computing methodologies~Artificial intelligence}

\keywords{Reasoning Patterns, AI Coding, Reasoning Trace Analysis, Reasoning Pattern Mining, Inference-Time Optimization}

\maketitle

\section{Introduction}

Large reasoning models (LRMs) are rapidly reshaping AI coding~\cite{codethinksurvey,sweagent2024}. Unlike conventional code language models that generate an answer directly, LRMs produce a sequence of intermediate reasoning steps before arriving at a solution. This capability is particularly valuable for coding tasks, where solving a problem often involves intermediate decisions, such as decomposing requirements and designing an algorithm for code generation or tracking program states during execution reasoning. The reasoning process is therefore exposed as part of the model's output rather than remaining implicit.

The availability of these reasoning traces has led to a growing body of research on their role in code-related tasks~\cite{wei2022cot,deepseekr1}. 
Existing studies show that intermediate reasoning can improve task performance, but also reveal substantial inefficiencies. Reasoning traces may contain incorrect assumptions, redundant exploration, repeated revisions, and prolonged deliberation that adds inference cost without improving the final result~\cite{revisitcotcodegen,cotcodequality,thinkingpatterns,overthinking2025}. These observations have motivated the use of reasoning traces for purposes beyond generating an answer, including self-correction~\cite{reflexion,selfrefine}, evaluation~\cite{codevisionary2025,judgeuncertainty2025}, and inference-time control~\cite{seer2025,turncontrol2025}. 

However, most existing studies examine reasoning from the perspective of individual tasks or specific behaviors. They focus on final task performance~\cite{codex,livecodebench,swebench2024}, prompting strategies~\cite{wei2022cot,selfconsistency,codecot}, or particular reasoning behaviors in tasks such as code generation~\cite{revisitcotcodegen,cotcodequality,thinkingpatterns,reasoningsteps2026}. This leaves a more basic question largely unanswered: do reasoning traces contain recurring structures that characterize how LRMs solve coding tasks? If such structures exist, it is also unclear which of them are associated with successful or unsuccessful reasoning, and whether they can be used to improve the efficiency and reliability of inference. Addressing these questions requires looking beyond individual reasoning traces and examining them as sequences of recurring reasoning actions.

To address these questions, we conduct an empirical study of reasoning patterns in AI coding. We collect reasoning traces generated by LRMs across four representative coding tasks: code generation, code execution reasoning, program debugging, and code translation. We transform these traces into structured reasoning action sequences using a unified action taxonomy. This representation allows us to analyze reasoning in terms of observable actions rather than the surface form of generated text. We then examine reasoning patterns at two levels. General patterns capture recurring task-level reasoning structures, whereas positive and negative patterns capture reasoning behaviors at the action level that distinguish valid from invalid trajectories. Finally, we investigate whether the discovered patterns can be leveraged for inference-time analysis and control of LRM behavior.

Specifically, we investigate the following research questions:

\begin{itemize}[itemsep=1.5mm]

\item \textbf{RQ1 (General Patterns): What task-level reasoning patterns emerge across different coding tasks?}

Different coding tasks impose distinct reasoning demands, yet it remains unclear whether these differences are reflected in recurring task-level structures. To characterize such structures, we represent reasoning traces as action sequences using a unified action taxonomy and mine recurring action fragments across the four coding tasks. We then examine the resulting patterns to identify the reasoning strategies that characterize each task.

\item \textbf{RQ2 (Positive and Negative Patterns): What action-level reasoning patterns distinguish valid from invalid reasoning trajectories?}

General patterns capture differences across coding tasks, but they do not reveal the local reasoning behaviors underlying trace validity. We therefore compare judge-labeled valid and invalid reasoning traces and mine recurring action fragments that characterize the two groups. We assess trace validity using an ensemble LLM-as-a-judge protocol and identify positive patterns prevalent in the valid-labeled group and negative patterns prevalent in the invalid-labeled group.

\item \textbf{RQ3 (Analysis and Control): Can the discovered reasoning patterns support diagnostic analysis and inference-time control of LRM behavior?}

We further examine how the discovered patterns can be used. Specifically, we evaluate a diagnostic task-identification probe and two inference-time controls. First, we use general-pattern features to identify coding tasks from reasoning traces. Second, we augment prompts with guidance derived from positive and negative patterns to steer reasoning behavior. Third, we use negative pattern-aware early stopping to terminate inference when such patterns are detected and quantify the resulting accuracy--token trade-off.

\end{itemize}

Our study yields three main findings. At the task level, coding tasks exhibit distinct reasoning structures: code generation follows a design-to-implementation flow, code execution reasoning centers on state tracing, debugging follows an evidence--fix--recheck loop, and code translation centers on semantic mapping. At the action level, judge-labeled valid and invalid trajectories exhibit distinct positive and negative patterns. Valid traces show verification-driven convergence, whereas invalid traces exhibit recurring behaviors such as repeated redesign, reflection, and knowledge recall without resolution. Beyond characterization, the discovered patterns support a diagnostic probe and two inference-time controls across the evaluated task--model settings. General-pattern features support diagnostic task identification, pattern-guided prompting yields higher mean pass@1, and negative pattern-aware early stopping reduces API-reported token usage with a corresponding decrease in pass@1. Each evaluation also includes previously unseen models.

In summary, this paper makes the following contributions:

\begin{itemize}[itemsep=1.2mm]

\item We establish a unified reasoning-action representation for cross-task analysis of LRM reasoning traces and conduct an empirical study across four representative AI coding tasks.

\item We characterize reasoning in AI coding at two levels, identifying general patterns at the task level and positive and negative patterns at the action level from judge-labeled valid and invalid reasoning trajectories, respectively.

\item We demonstrate the utility of these patterns for AI coding through a diagnostic task-identification probe and two inference-time controls: pattern-guided prompting and negative pattern-aware early stopping.

\end{itemize}

\section{Methodology}  

%\subsection{Study Overview}

\begin{figure}[t!]
\centering
\includegraphics[width=\linewidth]{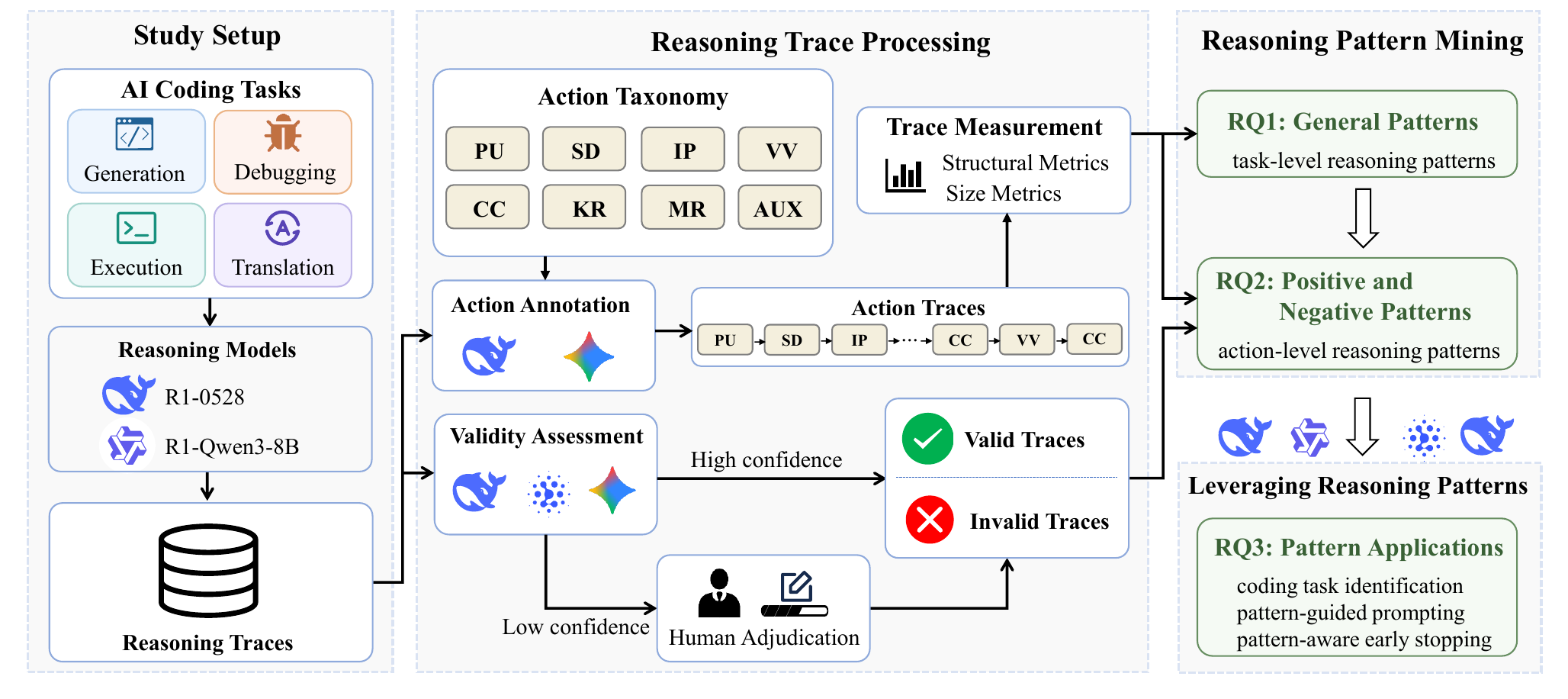}
\caption{Overview of the Empirical Study.}
\vspace{-2pt}
\label{fig:overview}
\end{figure}

Figure~\ref{fig:overview} presents an overview of our empirical study.
First, we collect reasoning traces generated by two LRMs across four representative AI coding tasks and assess the validity of each trace.
Second, we convert each trace into a sequence of reasoning actions using a unified reasoning-action taxonomy and quantify the resulting behaviors through size and structural metrics.
Third, we mine recurring patterns from the resulting action traces, including task-level general patterns for RQ1 and action-level positive and negative patterns for RQ2.
Finally, in RQ3, we evaluate how the discovered patterns can be used for AI coding through a diagnostic task-identification probe and two inference-time controls: prompting guided by positive and negative patterns, and early stopping guided by negative patterns.

\subsection{Study Setup}
\label{sec:study-setup}
\subsubsection{\textbf{Tasks and Datasets}}
To study reasoning patterns in AI coding across diverse problem-solving settings, we consider four representative coding tasks covering program synthesis, comprehension, maintenance, and transformation:

\begin{itemize}[itemsep=1.5mm]

\item \textit{Code generation} requires a model to synthesize executable programs from natural-language specifications.
We use LiveCodeBench v4--v6~\cite{livecodebench}, comprising 442 instances.

\item \textit{Code execution reasoning} requires a model to infer the output of a program through static reasoning rather than actual execution.
We use the input--output prediction subset of CodeSense~\cite{codesense}, comprising 382 instances.

\item \textit{Program debugging} requires a model to localize and repair defects in buggy programs.
We use DebugBench~\cite{debugbench}, comprising 500 instances.

\item \textit{Code translation} requires a model to transform programs across programming languages while preserving their semantics.
We use ClassEval-T~\cite{classeval-t}, comprising 376 instances.

\end{itemize}

All four tasks provide objective correctness oracles, which allow us to analyze both reasoning traces and task outcomes under a unified experimental setting.

\subsubsection{\textbf{Models}}
For RQ1 and RQ2, we use two publicly available LRMs, DeepSeek-R1-0528 (R1-0528) and its Qwen3-8B distilled variant, DeepSeek-R1-0528-Qwen3-8B (R1-Qwen3-8B)~\cite{deepseekr1}. Both models use identical prompts and decoding configurations, and we collect their complete reasoning traces and final responses for analysis. This pair holds the task and collection conditions fixed while exposing a within-lineage scale and distillation contrast; agreement between them is therefore interpreted as within-lineage consistency rather than independent model-family replication.
For RQ3, we additionally evaluate two previously unseen models, DeepSeek-V4-Flash (DS-V4-F)~\cite{deepseekv4} and GLM-5.1~\cite{glm51}, to test transfer of fixed pattern-derived guidance. DS-V4-F provides an additional model within the DeepSeek family, whereas GLM-5.1 provides one cross-family evaluation.

\subsection{Reasoning Trace Processing}
\label{sec:trace-processing}

\subsubsection{\textbf{Reasoning Trace Collection}}
For each task instance, we instantiate the benchmark-provided prompt template with the corresponding instance-specific input, such as a problem description, starter code, buggy program, or source-language snippet. We submit each instantiated prompt independently to both models in thinking mode under a unified decoding configuration (temperature \(=0.6\), top-\(p=0.95\)).

We extract the content enclosed by the \texttt{<think>} tags as the reasoning trace. To ensure completeness, we regenerate any response with an empty reasoning trace or lacking the information required for annotation or task evaluation, until a complete trace and task output are obtained.

After excluding incomplete, unparsable, or unlabeled responses, the resulting discovery dataset contains 3{,}376 reasoning traces across the four tasks and two models. Specifically, we retain 442/440 traces for code generation, 377/378 for code execution reasoning, 499/496 for program debugging, and 372/372 for code translation on R1-0528/R1-Qwen3-8B, respectively. %In total, R1-0528 contributes 1{,}690 traces and R1-Qwen3-8B contributes 1{,}686 traces.

\subsubsection{\textbf{Reasoning Trace Validity Assessment}}
To distinguish valid from invalid reasoning behaviors in RQ2, we assess the validity of each reasoning trace using an ensemble LLM-as-a-judge protocol~\cite{llmjudge,customizedevaluator}. Trace-level validity is distinct from task-level functional correctness: a trace may lead to a correct final answer while containing logical inconsistencies, unsupported transitions, or unnecessary reasoning detours. We consider a reasoning trace \emph{valid} if its reasoning is internally consistent and progresses coherently toward the final solution.

We employ three judge models, DeepSeek-V3.2~\cite{deepseekv32}, Gemini-3-Flash~\cite{gemini3flash}, and GLM-5.1~\cite{glm51}. We use a unified evaluation prompt that provides the same validity criteria and requires each judge to independently classify a trace as \emph{valid} or \emph{invalid} and report a confidence score in the range \([0,1]\). To reduce the effect of stochastic judgment variation, we query each judge three times for each trace, yielding nine judgments per trace.

We aggregate the nine judgments using confidence-weighted voting. For a candidate label \(\ell\), its normalized support is defined as:
\begin{equation}
S(\ell)=
\frac{\sum_{m\in\mathcal{M}}\sum_{r=1}^{3} c_{m,r}\cdot \mathbf{1}[y_{m,r}=\ell]}
{\sum_{m\in\mathcal{M}}\sum_{r=1}^{3} c_{m,r}}.
\end{equation}
We assign the label with the larger normalized support and use its support as the aggregate confidence. Traces with an aggregate confidence above 0.8 are automatically assigned the corresponding label; the remaining traces are adjudicated independently by two human annotators. The resulting labels define the valid and invalid reasoning-trace groups used in RQ2. The final dataset contains 2{,}452 valid and 924 invalid traces. R1-0528 contributes 1{,}439 valid and 251 invalid traces, while R1-Qwen3-8B contributes 1{,}013 valid and 673 invalid traces.

\subsubsection{\textbf{Reasoning Action Taxonomy and Annotation}}
Reasoning traces are free-form natural-language text, making systematic comparison across tasks and models difficult. To enable structured analysis, we represent each reasoning trace as a sequence of fine-grained \textit{reasoning actions}, where a reasoning action is defined as a contiguous text span expressing a distinct reasoning behavior.

Drawing on prior studies of LLM reasoning~\cite{codethinksurvey,thinkingpatterns,reflexion,selfrefine}, we define a taxonomy of eight reasoning actions that captures the major stages of code reasoning, ranging from task understanding and solution construction to verification and reflection. To strengthen its validity, the taxonomy is further inspected by two researchers with over five years of experience in code intelligence systems, who independently evaluated the coverage, granularity, and ambiguity of the action definitions and iteratively refined them to resolve overlaps. The resulting taxonomy provides a task-independent representation of reasoning behaviors across different coding tasks. Table~\ref{tab:taxonomy} summarizes the eight reasoning actions.

\begin{table}[t]
\centering
\caption{Taxonomy of Reasoning Actions}
\label{tab:taxonomy}
\papertablestyle
\begin{tabularx}{\linewidth}{>{\centering\arraybackslash}p{0.08\linewidth}>{\raggedright\arraybackslash}p{0.29\linewidth}>{\raggedright\arraybackslash}X}
\toprule
Abbr. & Reasoning Action & Description \\
\midrule
PU  & Problem understanding & Understanding and clarifying the problem requirements. \\
\rowcolor{blue!5}
SD  & Solution design & Designing the high-level solution approach and strategy. \\
IP  & Implementation planning & Planning the detailed implementation steps and logic. \\
\rowcolor{blue!5}
CC  & Code construction & Writing and refining the actual code. \\
VV  & Verification and validation & Testing, debugging, and validating the solution. \\
\rowcolor{blue!5}
KR  & Knowledge retrieval & Retrieving and applying relevant knowledge. \\
MR  & Meta-reasoning & Reflecting on and monitoring the reasoning process itself. \\
\rowcolor{blue!5}
AUX & Auxiliary & Auxiliary statements that support the reasoning flow. \\
\bottomrule
\end{tabularx}
\end{table}

Using this taxonomy, we automatically annotate each reasoning trace with DeepSeek-V3.2~\cite{deepseekv32} and Gemini-2.5 Pro~\cite{gemini25}. Specifically, each reasoning trace is segmented into contiguous spans, and each span is assigned exactly one reasoning action. To ensure consistent annotations, we require the models to generate structured JSON outputs containing the assigned reasoning actions, corresponding text spans, and their ordering. The annotation prompt specifies (1) the definitions of the eight reasoning actions with representative examples, (2) segmentation guidelines covering boundary criteria, granularity, and special-case handling, (3) priority rules for resolving ambiguous cases, and (4) the required JSON output schema. The complete annotation prompt is included in our replication package.

To assess annotation reliability, two human annotators independently re-annotated a randomly sampled subset of reasoning traces. Their annotations achieved substantial agreement with the LLM-generated annotations (Cohen's $\kappa=0.816$)~\cite{cohen1960kappa}, indicating that the proposed taxonomy and annotation protocol produce reliable reasoning-action annotations.

\subsubsection{\textbf{Action Trace Construction}}
The annotated reasoning actions are converted into an \textit{action trace}, defined as an ordered sequence of reasoning actions:
\begin{equation}
\tau=(s_1,s_2,\ldots,s_n),
\label{eq:action-trace}
\end{equation}
where each $s_i \in \mathcal{C}=\{PU,SD,IP,CC,VV,KR,MR,$ $AUX\}$ denotes one reasoning action from the taxonomy. We use the sequence in Eq.~\eqref{eq:action-trace} as the canonical representation for all subsequent analyses.

To emphasize transitions between reasoning stages while reducing the influence of repeated elaboration within the same action, we further derive a \textit{collapsed action trace} $\bar{\tau}$ by merging adjacent identical actions. For example,
\[
(SD,SD,KR,SD,VV,VV)
\twoheadrightarrow
(SD,KR,SD,VV).
\]

\subsubsection{\textbf{Action Trace Measurement}}
\label{sec:structural-metrics}
We characterize each reasoning action trace using two groups of metrics: \emph{size metrics}, which quantify the length and output volume of a reasoning trace, and \emph{structural metrics}, which capture the distribution and transition dynamics of reasoning actions.

\textbf{Size metrics} quantify the length of the reasoning trace at both the action and token levels. We report the raw action-trace length (n) and collapsed length (m), with their ratio (n/m) measuring the degree of consecutive action repetition. We additionally compute the number of reasoning tokens generated by the model using a consistent tokenizer across all comparisons. These tokenizer-based counts characterize generated reasoning volume and are distinct from API-reported token usage.

\textbf{Structural metrics} characterize the composition of reasoning actions and their transition dynamics. For each action \(c\), its \emph{allocation} is defined as the fraction of action spans in a trace assigned to \(c\). To characterize reasoning flow, we construct a transition matrix \(P\), where \(P_{ab}\) denotes the probability of transitioning from action \(a\) to action \(b\). We estimate \(P_{ab}\) by counting adjacent \(a\rightarrow b\) transitions in collapsed traces and normalizing by the total number of outgoing transitions from \(a\). We then quantify transition diversity using the stationary-distribution-weighted average of the row-wise transition entropies:
\begin{equation}
H(P)=\sum_{a}\pi_a\left(-\sum_{b}P_{ab}\log_2P_{ab}\right),
\end{equation}
where \(\pi\) denotes the stationary distribution of \(P\). Higher entropy indicates greater diversity in the actions that can follow a given action, reflecting a less concentrated transition structure.

\subsection{Reasoning Pattern Mining}
\label{sec:pattern-mining}

This section describes how we mine reasoning patterns from the resulting reasoning action traces and characterize them at two granularities: \textit{general patterns} at the task level and \textit{positive and negative patterns} at the action level, corresponding to RQ1 and RQ2, respectively.

\subsubsection{\textbf{Reasoning Pattern Definition}}

A \textit{reasoning pattern} is a recurring action-fragment signature identified from a collection of LRM reasoning action traces. Given a trace set \(\mathcal{T}\), the \emph{support} of a signature \(s\) is the fraction of traces in \(\mathcal{T}\) whose collapsed action trace contains \(s\):
\begin{equation}
\operatorname{supp}(s,\mathcal{T}) =
\frac{|{t\in\mathcal{T}: s\subseteq t}|}{|\mathcal{T}|}.
\end{equation}
A signature is considered a candidate pattern when its support exceeds a predefined minimum support threshold. We treat reasoning patterns as observable regularities in action traces rather than as direct evidence of the model's latent cognitive mechanisms.

We distinguish two types of reasoning patterns according to their analysis granularity. \emph{General patterns} capture task-level reasoning structures through ordering relations among actions. We denote \(A\Rightarrow B\) when action \(A\) precedes action \(B\) in a trace, allowing arbitrary intermediate actions between them. Thus, this representation captures coarse-grained ordering relations without requiring the actions to be adjacent. 
\emph{Positive and negative patterns}, in contrast, capture local reasoning structures through adjacent action transitions. We denote \(A\rightarrow B\) when \(A\) and \(B\) occur consecutively, and \(A\leftrightarrow B\) when \(A\) and \(B\) repeatedly alternate over adjacent positions, forming a short cyclic loop. These representations provide the basis for identifying task-level regularities and distinguishing local patterns prevalent in valid and invalid reasoning trajectories, respectively.

\subsubsection{\textbf{Mining General Patterns}}
\label{sec:general-pattern-discovery}

General patterns are mined to identify recurring action structures that distinguish one coding task from the others. For each task, we extract candidate action-fragment signatures from its collapsed reasoning traces. We consider two forms of signatures: contiguous action \(n\)-grams with lengths of 4--5 and non-contiguous ordered action paths. For example, \(PU\Rightarrow SD\Rightarrow IP\Rightarrow CC\) denotes an ordered pattern in which the four actions occur in the specified order, while allowing arbitrary intermediate actions.

We then assess the task specificity of each candidate signature using its within-task support and its support difference from the other tasks. A signature is retained as a general pattern for a task only if its support within that task exceeds the minimum support threshold and its support advantage over the other tasks exceeds a predefined margin. The resulting general patterns capture recurring task-level reasoning structures and are subsequently used for task identification in RQ3.

\subsubsection{\textbf{Mining Positive and Negative Patterns}}
\label{sec:pattern-discovery}

Positive and negative patterns are mined to identify recurring action-level structures that distinguish valid reasoning traces from invalid ones. We examine the two trace groups from complementary perspectives: aggregate action allocation and local action transitions. The former characterizes how reasoning actions are distributed across a trace, whereas the latter captures recurring local structures in the action sequence.

The first perspective focuses on \emph{action allocation}. We define the \emph{closure share} as the combined proportion of action spans labeled verification and validation (\(VV\)), code construction (\(CC\)), or auxiliary (\(AUX\)). This aggregate measure captures the extent to which a trace allocates reasoning toward convergence-oriented actions. Here, \(AUX\) retains its taxonomy definition of auxiliary statements and does not specifically denote answer finalization. Conversely, the \emph{drift share} is the combined proportion of spans labeled solution design, meta-reasoning, or knowledge retrieval, capturing allocation toward actions that revisit, extend, or redirect the reasoning process. We summarize the relative balance between the two using the \emph{closure-minus-drift} score, defined as the closure share minus the drift share. Positive values indicate greater allocation toward closure-oriented actions, whereas negative values indicate greater allocation toward drift-oriented actions.

The second perspective focuses on \emph{local structural patterns}. From collapsed action traces, we extract local structural patterns represented by adjacent action transitions (\(A\rightarrow B\)) and short alternating loops (\(A\leftrightarrow B\)). Unlike the task-level ordered paths used for general patterns, these structures capture local reasoning dynamics that occur over neighboring actions. These patterns capture recurring action-level structures that are not reflected in aggregate allocation measures. Whereas closure and drift shares characterize the overall allocation of reasoning actions, local patterns identify the specific action transitions and short sequences through which these allocation tendencies manifest.

For local patterns, we identify group-enriched candidates based on both support and support contrast. A candidate pattern must exceed a minimum support threshold in its enriched group and exhibit a minimum support difference relative to the comparison group. Patterns more prevalent in valid traces are labeled \emph{positive}, whereas those more prevalent in invalid traces are labeled \emph{negative}. We assess the statistical significance of the resulting group differences using two-sided Fisher's exact tests with Benjamini--Hochberg correction~\cite{benjamini1995fdr}.

\subsection{Leveraging Reasoning Patterns}
\label{sec:pattern-applications}

\subsubsection{\textbf{Coding Task Identification}}
We train a supervised probe on features extracted from reasoning action traces to identify the coding task performed by an LRM. The features operationalize the general patterns identified in RQ1 by encoding how a trace allocates actions, transitions between them, and distributes them across reasoning stages.
We construct 57 features in seven groups: (1) eight action-allocation features; (2) two trace-size features; (3) twelve transition-probability features; (4) three repetition features; (5) twenty-four position-aware action-allocation features; (6) four action-effort features; and (7) four aggregate features. The probe uses only reasoning-trace features and has no access to task descriptions, final answers, or external metadata.

\subsubsection{\textbf{Pattern-Guided Prompting}}
We augment the original task prompts with guidance derived from the positive and negative patterns identified in RQ2. Positive patterns are presented as desirable reasoning structures to encourage during inference, whereas negative patterns are formulated as constraints that discourage recurring undesirable transitions or loops, such as \(SD{\rightarrow}KR{\rightarrow}SD\) and \(VV{\rightarrow}SD{\rightarrow}MR\). The original task instruction is kept unchanged, and the pattern-derived guidance is appended as an additional inference-time signal, allowing us to examine whether empirically identified reasoning patterns can influence downstream coding behavior without modifying the underlying model.

\subsubsection{\textbf{Negative Pattern-Aware Early Stopping}}

We further use negative patterns as online signals to terminate unproductive reasoning before the model completes its original generation. Because full-trace LLM annotation is unsuitable for online intervention, we employ a lightweight rule-based action mapper separate from the LLM-based annotation used for pattern discovery in RQ1 and RQ2. The mapper processes streamed \texttt{reasoning\_content}, segments it into micro-units, and assigns each unit one of eight fine-grained actions and one of four coarse states (Explore, Build, Verify, or Wrap) using lexical cues. Consecutive units are grouped into segments based on coarse-state changes, fine-action transitions, and lexical novelty, and each segment receives its majority action label. Adjacent segments with identical labels are collapsed before matching task-specific motifs. The mapper operates locally without additional model or API calls.

At each completed segment, the stopping policy combines negative motifs with recent exploration, lexical novelty, and build/verify history. A loop trigger requires sustained drift and insufficient closure evidence, subject to task-specific minimum-progress guards, while a separate closure trigger requires strong closure evidence and answer or wrap indicators. A repeated-text detector provides an additional safeguard against degenerate output. When a stopping condition is met, generation is terminated and the original task with the generated reasoning prefix is submitted to the same model in a second request for finalization. The request uses answer-only instructions and disables thinking where supported; any reasoning generated during finalization remains included in the API-reported token usage.

\section{RQ1: General Patterns}
\label{sec:rq1}

\subsection{Setup}
RQ1 characterizes task-level reasoning behaviors using the general pattern mining procedure described in Section~\ref{sec:general-pattern-discovery}. We analyze 3{,}376 collapsed reasoning action traces generated by two LRMs across four coding tasks. A general pattern is retained if its within-task support exceeds 0.05 and its support difference from the other tasks exceeds 0.03. These thresholds were calibrated by two experts through empirical inspection of preliminary results.

\subsection{Results}

We identify one task-specific general reasoning pattern for each coding task. Table~\ref{tab:rq1-patterns} reports representative signatures and within- versus cross-task support; the large gaps (e.g., $91.4\%$ vs.\ $17.0\%$ for code generation and $45.9\%$ vs.\ $2.2\%$ for execution reasoning on R1-0528) confirm their task specificity.

\begin{table*}[t]
\centering
\caption{General Reasoning Patterns Identified across Four Coding Tasks}
\label{tab:rq1-patterns}
\footnotesize
\setlength{\tabcolsep}{8pt}
\renewcommand{\arraystretch}{1.0}
\resizebox{\textwidth}{!}{%
\begin{tabular}{lllll}
\toprule
\multirow{2}{*}{Task}&\multirow{2}{*}{General Pattern} & \multirow{2}{*}{Representative Signature} & \multicolumn{2}{c}{Support (within-task vs.\ cross-task)} \\
\cmidrule(lr){4-5}
 &  &  & R1-0528 & R1-Qwen3-8B \\
\midrule
Generation & Design-to-implementation & \(PU\Rightarrow SD\Rightarrow IP\Rightarrow VV\) & 91.4\% vs.\ 17.0\% & 88.0\% vs.\ 22.0\% \\
\rowcolor{blue!5}
Execution Reasoning & State tracing & \(PU\Rightarrow\{VV,MR\}^{*}\) & 45.9\% vs.\ 2.2\% & 53.5\% vs.\ 3.4\% \\
Debugging & Evidence-fix-recheck & \(\{SD,IP\}\Rightarrow CC\Rightarrow VV\) & 68.1\% vs.\ 21.5\% & 70.8\% vs.\ 20.4\% \\
\rowcolor{blue!5}
Translation & Semantic mapping-to-implementation & \(PU\Rightarrow SD\Rightarrow IP\Rightarrow CC\) & 36.5\% vs.\ 9.5\% & 30.1\% vs.\ 16.2\% \\
\bottomrule
\end{tabular}
}
\end{table*}

% \begin{table*}[t]
% \centering
% \caption{Macro Reasoning Patterns across Four Coding Tasks}
% \label{tab:rq1-patterns}
% \footnotesize
% \setlength{\tabcolsep}{8pt}
% \renewcommand{\arraystretch}{1.3}
% \resizebox{\textwidth}{!}{%
% \begin{tabular}{llllll}
% \toprule
% \multirow{2}{*}{Task}&\multirow{2}{*}{Macro Pattern} & \multirow{2}{*}{Representative Signature} & \multicolumn{3}{c}{Support (within-task vs.\ cross-task)} \\
% \cmidrule(lr){4-6}
%  &  &  & R1-0528 & R1-Qwen3-8B & \textcolor{blue}{GLM-5.1} \\
% \midrule
% Generation & Design-to-implementation & \(PU\Rightarrow SD\Rightarrow IP\Rightarrow VV\) & 91.4\% vs.\ 17.0\% & 88.0\% vs.\ 22.0\% & \textcolor{blue}{66.7\% vs.\ 10.5\%} \\
% \rowcolor{blue!5}
% Execution Reasoning & State tracing & \(PU\Rightarrow\{VV,MR\}^{*}\) & 45.9\% vs.\ 2.2\% & 53.5\% vs.\ 3.4\% & \textcolor{blue}{54.3\% vs.\ 17.1\%} \\
% Debugging & Evidence-fix-recheck & \(\{SD,IP\}\Rightarrow CC\Rightarrow VV\) & 68.1\% vs.\ 21.5\% & 70.8\% vs.\ 20.4\% & \textcolor{blue}{70.9\%$^{\dagger}$ vs.\ 32.4\%} \\
% \rowcolor{blue!5}
% Translation & Semantic mapping-to-implementation & \(PU\Rightarrow SD\Rightarrow IP\Rightarrow CC\) & 36.5\% vs.\ 9.5\% & 30.1\% vs.\ 16.2\% & \textcolor{blue}{51.7\%$^{\dagger}$ vs.\ 21.0\%} \\
% \bottomrule
% \end{tabular}
% }
% 
% \vspace{2pt}
% \parbox{0.9\textwidth}{\footnotesize
% \textcolor{blue}{$^{\dagger}$ For GLM-5.1, the most discriminative signatures are \(PU\Rightarrow VV\Rightarrow CC\Rightarrow VV\) for debugging and \(PU\Rightarrow SD\Rightarrow CC\Rightarrow VV\) for translation.}}
% \end{table*}

\textbf{Code Generation Follows a Design-to-Implementation Pattern.}
Both models progress from solution design through implementation planning to verification. R1-0528 interleaves design, planning, and verification (\(PU\rightarrow SD\rightarrow VV\rightarrow SD\rightarrow IP\rightarrow VV\)), with \(IP\leftrightarrow VV\) dominant; R1-Qwen3-8B follows the same structure while incorporating additional meta-reasoning (\(PU\rightarrow SD\rightarrow MR\rightarrow SD\rightarrow VV\)).

\textbf{Code Execution Reasoning Follows a State-Tracing Pattern.}
Both models simulate program states and verify intermediate results. R1-0528 follows a compact trajectory of state tracing and output verification (\(PU\rightarrow VV\rightarrow MR\)) with the lowest transition entropy across the four tasks (1.690), whereas R1-Qwen3-8B exhibits a longer verification--reflection cycle (\(PU\rightarrow VV\rightarrow MR\rightarrow VV\)), indicating more iterative transitions within the same structure.

\textbf{Program Debugging Follows an Evidence-Fix-Recheck Pattern.}
Both models alternate between fault validation, code modification, and re-verification. R1-0528 enters verification early and repeatedly alternates \(IP\) and \(VV\) (\(PU\rightarrow VV\rightarrow IP\rightarrow VV\rightarrow CC\rightarrow VV\)), with frequent \(PU\rightarrow VV\) and \(CC\rightarrow VV\) transitions; R1-Qwen3-8B follows the same cycle but revisits \(SD\) more often (\(PU\rightarrow SD\rightarrow VV\rightarrow SD\rightarrow IP\rightarrow CC\rightarrow VV\rightarrow CC\)).

\textbf{Code Translation Follows a Semantic Mapping Pattern.}
Both models establish source--target correspondences before implementation. R1-0528 transitions directly from semantic mapping to implementation and validation (\(PU\rightarrow SD\rightarrow IP\rightarrow VV\rightarrow IP\)), whereas R1-Qwen3-8B revisits \(SD\) more often (\(PU\rightarrow SD\rightarrow IP\rightarrow SD\rightarrow IP\)); the shared signature \(PU\Rightarrow SD\Rightarrow IP\Rightarrow CC\) captures semantic reconstruction followed by code construction.

\textbf{Shared Reasoning Scaffold.}
Despite different objectives, all tasks share an understand-then-act scaffold. \(PU\) appears in more than 95\% of traces, after which traces diverge into task-specific sequences involving \(SD\), \(IP\), \(VV\), and \(CC\). \(VV\) is prominent in generation, execution reasoning, and debugging, while \(IP\) is prominent in generation and translation. The raw-to-collapsed length ratio remains stable (1.3--1.7), indicating comparable consecutive repetition. Thus, task-specific structure emerges primarily after the shared problem-understanding stage.

\textbf{Cross-Model Comparison.}
The two R1 models exhibit highly consistent task-level structures but differ in realization: R1-Qwen3-8B generates longer traces than R1-0528 (+37.1\% for code generation, +133.1\% for execution reasoning, +21.8\% for debugging, and +7.6\% for translation) and exhibits equal or higher transition entropy, particularly for execution reasoning and debugging. Thus, the models share task-level structures, while R1-Qwen3-8B tends to realize them through longer and more dispersed trajectories.

\begin{tcolorbox}[width=\linewidth, boxrule=0.8pt, left=2pt, right=2pt, top=2pt, bottom=2pt, colback=gray!10, colframe=black]
\textbf{Finding 1.}
Each coding task exhibits a distinct general reasoning pattern that reflects its underlying problem-solving strategy. Code generation follows a design-to-implementation pattern, code execution reasoning relies on state tracing, program debugging follows an evidence-fix-recheck cycle, and code translation is driven by semantic mapping followed by implementation.
\end{tcolorbox}

\section{RQ2: Positive and Negative Patterns}
\label{sec:rq2}

\subsection{Setup}
RQ2 characterizes action-level reasoning differences between valid and invalid reasoning traces using the positive and negative pattern mining procedure described in Section~\ref{sec:pattern-discovery}. We analyze 2{,}452 valid and 924 invalid traces generated by the two primary R1 models across four coding tasks. A pattern is retained when its support in the enriched group exceeds 0.08 and its support difference from the comparison group exceeds 0.06. These thresholds were calibrated by two domain experts through empirical inspection of preliminary results. %They serve as an effect-size filter prior to statistical testing: the support threshold removes sparse occurrences (requiring at least two traces even in the smallest subgroup), while the margin constraint filters patterns with negligible separation between valid and invalid groups.

\subsection{Results}

We compare valid and invalid traces through aggregate action allocation and local transition patterns, revealing fine-grained differences in action dynamics.

\begin{table}[t]
\centering
\caption{Length and Action Allocation of Valid and Invalid Reasoning Traces}
\label{tab:rq2-trace-stats}
\footnotesize
\setlength{\tabcolsep}{6pt}
\renewcommand{\arraystretch}{1.05}
\begin{tabular}{l rr rr}
\toprule
\multirow{2}{*}{Metric} & \multicolumn{2}{c}{R1-0528} & \multicolumn{2}{c}{R1-Qwen3-8B} \\
\cmidrule(lr){2-3}\cmidrule(lr){4-5}
 & Valid & Invalid & Valid & Invalid \\
\midrule
Reasoning tokens & 3{,}690 & 6{,}609 & 4{,}330 & 6{,}991 \\
Raw action length & 18.58 & 24.96 & 22.64 & 32.21 \\
Collapsed action length & 11.78 & 14.91 & 15.97 & 21.99 \\
\midrule
Closure share & 0.414 & 0.241 & 0.330 & 0.219 \\
Drift share & 0.247 & 0.391 & 0.313 & 0.464 \\
Closure$-$Drift & $+$0.167 & $-$0.150 & $+$0.017 & $-$0.245 \\
\bottomrule
\end{tabular}
\end{table}

\textbf{Invalid Traces Shift Action Allocation Toward Drift-oriented Actions.}
Table~\ref{tab:rq2-trace-stats} summarizes reasoning-token counts, action-trace lengths, and action allocation across the four oracle-backed tasks. Invalid traces are longer than valid traces in reasoning tokens, raw action length, and collapsed action length, indicating greater reasoning volume and more action steps.

The two groups also differ substantially in action allocation. Invalid traces exhibit lower closure shares (R1-0528: \(0.414 \rightarrow 0.241\); R1-Qwen3-8B: \(0.330 \rightarrow 0.219\)) and higher drift shares (R1-0528: \(0.247 \rightarrow 0.391\); R1-Qwen3-8B: \(0.313 \rightarrow 0.464\)). This shift corresponds to less allocation to verification, code construction, and auxiliary actions and greater allocation to solution design, meta-reasoning, and knowledge retrieval.

The closure-minus-drift score further summarizes this difference. Its value changes from positive to negative between valid and invalid traces for both models (R1-0528: \(+0.167 \rightarrow -0.150\); R1-Qwen3-8B: \(+0.017 \rightarrow -0.245\)), indicating that invalid traces allocate relatively more reasoning to drift-oriented than closure-oriented actions.

\begin{table}[t]
\centering
\caption{Positive and Negative Patterns Identified across Four Coding Tasks. \textcolor{gray}{Gray}-shaded cells denote negative patterns.}
\label{tab:rq2-pattern-catalog}
\papertablestyle
\begin{threeparttable}
\setlength{\tabcolsep}{8pt}
\renewcommand{\arraystretch}{1.0}
\begin{tabular}{llll}
\toprule
Task & Model & Positive/Negative Pattern & Representative Signature\\
\midrule

\multirow{8}{*}{Generation}
& \multirow{4}{*}{R1-0528} & Plan-verify-close &IP$\rightarrow$VV$\rightarrow$AUX\\
&  & Plan-verify iteration &IP$\rightarrow$VV$\rightarrow$IP$\rightarrow$VV$\rightarrow$IP\\
& & \cellcolor{gray!15}Knowledge-verify-reflect detour &\cellcolor{gray!15}KR$\rightarrow$VV$\rightarrow$MR\\
& & \cellcolor{gray!15}Design-knowledge detour &\cellcolor{gray!15}SD$\rightarrow$KR$\rightarrow$IP\\

\cmidrule(lr){2-4}

& \multirow{4}{*}{R1-Qwen3-8B} & Verify-plan-reverify &VV$\rightarrow$IP$\rightarrow$VV\\
& & Extended verify-plan &VV$\rightarrow$IP$\rightarrow$VV$\rightarrow$IP\\
& & \cellcolor{gray!15}Knowledge-design loop &\cellcolor{gray!15}KR$\leftrightarrow$SD\\
& & \cellcolor{gray!15}Reflection-knowledge loop &\cellcolor{gray!15}MR$\leftrightarrow$KR\\

% \cmidrule(lr){2-4}
% 
% & \multirow{2}{*}{\textcolor{blue}{GLM-5.1}} & \textcolor{blue}{Understand-design-plan} &\textcolor{blue}{PU$\rightarrow$SD$\rightarrow$IP}\\
% & & \cellcolor{gray!15}\textcolor{blue}{Reflection-design reopening} &\cellcolor{gray!15}\textcolor{blue}{MR$\rightarrow$SD$\rightarrow$VV$\rightarrow$MR}\\
% 
\midrule

\multirow{4}{*}{\makecell[l]{Execution\\Reasoning}}
& \multirow{2}{*}{R1-0528} & State-trace verification &PU$\rightarrow$VV$\rightarrow$MR$\rightarrow$VV\\
& & \cellcolor{gray!15}Understand-reopening reflection &\cellcolor{gray!15}MR$\rightarrow$PU$\rightarrow$MR\\

\cmidrule(lr){2-4}

& \multirow{2}{*}{R1-Qwen3-8B} &
State-trace verification &PU$\rightarrow$VV$\rightarrow$MR$\rightarrow$VV\\
& & \cellcolor{gray!15}Reflection-design reopening &\cellcolor{gray!15}MR$\rightarrow$SD$\rightarrow$\{PU,VV\}\\

% \cmidrule(lr){2-4}
% 
% & \multirow{2}{*}{\textcolor{blue}{GLM-5.1}} &
% \textcolor{blue}{State-trace rechecking} &\textcolor{blue}{PU$\rightarrow$VV$\rightarrow$MR$\rightarrow$VV}\\
% & & \cellcolor{gray!15}\textcolor{blue}{Reflection-knowledge detour} &\cellcolor{gray!15}\textcolor{blue}{MR$\rightarrow$KR$\rightarrow$VV}\\
% 
\midrule

\multirow{5}{*}{Debugging}
& \multirow{3}{*}{R1-0528} &
Change-verify-close & CC$\rightarrow$VV$\rightarrow$AUX\\
& & \cellcolor{gray!15}Verify-design-reflect reopening &\cellcolor{gray!15}VV$\rightarrow$SD$\rightarrow$MR\\
& & \cellcolor{gray!15}Design-reflect detour &\cellcolor{gray!15}SD$\rightarrow$MR$\rightarrow$IP\\

\cmidrule(lr){2-4}

& \multirow{2}{*}{R1-Qwen3-8B} &
Verify-plan-reverify & VV$\rightarrow$IP$\rightarrow$VV\\
& & \cellcolor{gray!15}Knowledge-design reopening &\cellcolor{gray!15}SD$\rightarrow$KR$\rightarrow$SD\\

% \cmidrule(lr){2-4}
% 
% & \multirow{2}{*}{\textcolor{blue}{GLM-5.1}} &
% \textcolor{blue}{Reflect-verify-repair} &\textcolor{blue}{MR$\rightarrow$VV$\rightarrow$CC}\\
% & & \cellcolor{gray!15}\textcolor{blue}{Reflect-change premature close} &\cellcolor{gray!15}\textcolor{blue}{MR$\rightarrow$CC$\rightarrow$VV$\rightarrow$AUX}\\
% 
\midrule

\multirow{6}{*}{Translation}
& \multirow{3}{*}{R1-0528} &
Design-plan-construct &SD$\rightarrow$IP$\rightarrow$CC\\
& & \cellcolor{gray!15}Design-reflect reopening &\cellcolor{gray!15}SD$\rightarrow$MR$\rightarrow$SD$\rightarrow$MR$\rightarrow$SD\\
& & \cellcolor{gray!15}Verify-plan-construct detour &\cellcolor{gray!15}VV$\rightarrow$IP$\rightarrow$CC\\

\cmidrule(lr){2-4}

& \multirow{3}{*}{R1-Qwen3-8B} &
Understand-design-plan-reflect & PU$\rightarrow$SD$\rightarrow$IP$\rightarrow$MR\\
& & \cellcolor{gray!15}Design-reflect replanning &\cellcolor{gray!15}SD$\rightarrow$MR$\rightarrow$SD$\rightarrow$IP\\
& & \cellcolor{gray!15}Understand-design reopening &\cellcolor{gray!15}PU$\rightarrow$SD$\rightarrow$PU$\rightarrow$SD\\

% \cmidrule(lr){2-4}
% 
% & \multirow{3}{*}{\textcolor{blue}{GLM-5.1}} &
% \textcolor{blue}{Verify-construct-reverify} &\textcolor{blue}{VV$\rightarrow$CC$\rightarrow$VV}\\
% & & \cellcolor{gray!15}\textcolor{blue}{Plan-reflect-replan} &\cellcolor{gray!15}\textcolor{blue}{IP$\rightarrow$MR$\rightarrow$IP}\\
% & & \cellcolor{gray!15}\textcolor{blue}{Construct-reflect loop} &\cellcolor{gray!15}\textcolor{blue}{CC$\leftrightarrow$MR}\\
% 
\bottomrule
\end{tabular}

%\vspace{2pt}
%\begin{tablenotes}[flushleft]
%\item[] $^{*}$ The \textcolor{gray}{gray} shaded cells are anti-patterns.
%\end{tablenotes}
\end{threeparttable}
\end{table}

%\FloatBarrier

\begin{figure*}[t!]
\centering
\includegraphics[width=\textwidth]{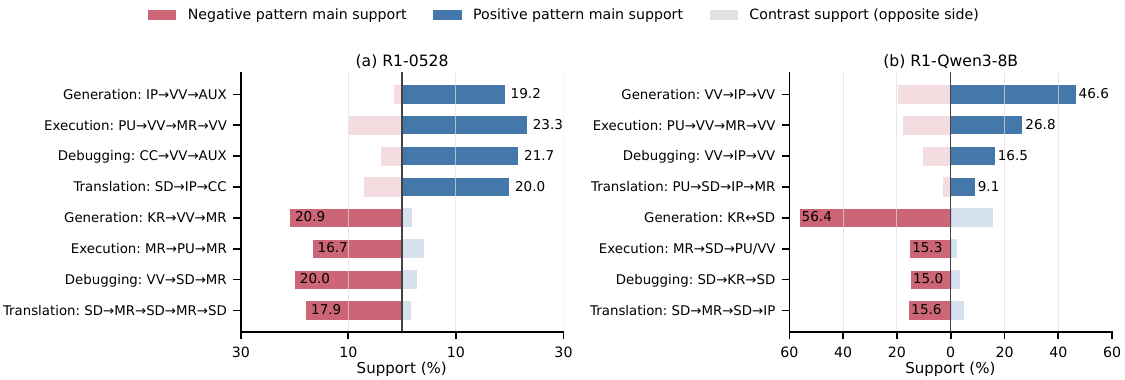}
\caption{Support Comparison between Positive and Negative Patterns.}
\label{fig:rq2-patterns}
\end{figure*}

\textbf{Positive and Negative Patterns Distinguish Convergent Iteration from Repeated Reopening.}
We next examine whether the differences in action allocation are reflected in local reasoning structures. Table~\ref{tab:rq2-pattern-catalog} and Figure~\ref{fig:rq2-patterns} summarize the local positive and negative patterns and their support contrasts across task--model settings.
Valid and invalid traces exhibit distinct local transition patterns. For example, for R1-Qwen3-8B on code generation, the positive pattern \(VV \rightarrow IP \rightarrow VV\) occurs in 46.6\% of valid traces, compared with 19.6\% of invalid traces. In contrast, the negative pattern \(KR \leftrightarrow SD\) occurs in 56.4\% of invalid traces but only 15.9\% of valid traces. A similar contrast appears in debugging for R1-0528: \(CC \rightarrow VV \rightarrow AUX\) occurs in 21.7\% of valid traces versus 4.0\% of invalid traces, whereas \(VV \rightarrow SD \rightarrow MR\) occurs in 20.0\% of invalid traces versus 2.7\% of valid traces.

These patterns indicate two qualitatively different forms of iteration. Positive patterns return to verification after planning or implementation, such as \(VV \rightarrow IP \rightarrow VV\), or proceed from implementation to verification and auxiliary completion, such as \(CC \rightarrow VV \rightarrow AUX\). Negative patterns, in contrast, more often repeat transitions among design, reflection, and knowledge retrieval, including cycles such as \(KR \leftrightarrow SD\). Thus, the distinction between valid and invalid traces is reflected not only in overall action allocation but also in how actions are locally sequenced.

We assess representative support contrasts using two-sided Fisher's exact tests with Benjamini--Hochberg correction; complete support, count, \(p\)-value, and \(q\)-value results are provided in the replication package. The association with validity depends on the actions and transition structure, not iteration frequency alone.

\textbf{Cross-model Consistency with Task-dependent Strength.}
Finally, we examine the consistency of the observed validity contrasts across models and coding tasks. The distinction between closure-oriented and drift-oriented reasoning is observed in both primary R1 models, suggesting that the identified differences are not specific to a single model within this model family. The same direction of contrast is also reflected in the local patterns: valid traces more frequently contain verification-centered transitions, whereas invalid traces more frequently contain repeated transitions involving design revision, reflection, and knowledge retrieval.

The magnitude of these differences varies across settings. For R1-Qwen3-8B, invalid traces exhibit higher drift shares and more frequent cycles involving \(SD\), \(MR\), and \(KR\). This difference is also reflected in trace length. For code generation, invalid traces contain substantially more reasoning tokens than valid traces for both R1-0528 (17,163 vs. 7,102) and R1-Qwen3-8B (14,925 vs. 8,598), with a stronger concentration of drift-oriented actions. Code execution reasoning and debugging exhibit similar but more structured contrasts, whereas the differences are smaller and less symmetric for code translation.

\begin{tcolorbox}[width=\linewidth, boxrule=0.8pt, left=2pt, right=2pt, top=2pt, bottom=2pt, colback=gray!10, colframe=black]
\textbf{Finding 2.}
Valid and invalid reasoning traces differ in trace length, action allocation, and local reasoning patterns. Positive patterns in valid traces predominantly reflect verification-centered iteration, whereas negative patterns in invalid traces more often involve repeated cycles of design revision, reflection, and knowledge retrieval.
\end{tcolorbox}

\section{RQ3: Leveraging Reasoning Patterns}
\label{sec:rq3}

To examine how the discovered reasoning patterns can be leveraged at inference time, we evaluate three applications covering task identification, pattern-guided prompting, and pattern-aware early stopping. Specifically, we investigate:

\begin{itemize}
    \item \textbf{RQ3.1} \textit{Can general reasoning patterns support reliable identification of different AI coding tasks?  (Section \ref{sec:RQ3.1})}

    \item \textbf{RQ3.2} \textit{How does prompting guided by positive and negative patterns affect coding performance? (Section \ref{sec:RQ3.2})}

    \item \textbf{RQ3.3} \textit{How does negative pattern-aware early stopping affect coding accuracy and token usage? (Section \ref{sec:RQ3.3})}
\end{itemize}

\subsection{Task Identification Using General-Pattern Features}
\label{sec:RQ3.1}

\textbf{Setup.}
Following the task-identification probe described in Section~\ref{sec:pattern-applications}, RQ3.1 examines whether coding tasks can be identified using general-pattern features derived solely from action traces. The evaluation dataset contains 3{,}376 reasoning traces generated by two models across four coding tasks. We evaluate two standard classifiers, logistic regression and random forest, using 5-fold stratified cross-validation. We report classification accuracy as the mean and standard deviation across folds; for the four-class setting, random guessing yields an accuracy of 25\%.

To evaluate cross-model transfer, we train the classifiers on R1-0528 traces and directly apply them to reasoning traces generated by the two previously unseen models, DS-V4-F and GLM-5.1. The target-model traces are used only for testing, with no retraining or model-specific calibration. We extract the same 57 action-trace features and predict the same four task labels.

\begin{figure}[t!]
\centering
\includegraphics[width=0.75\linewidth]{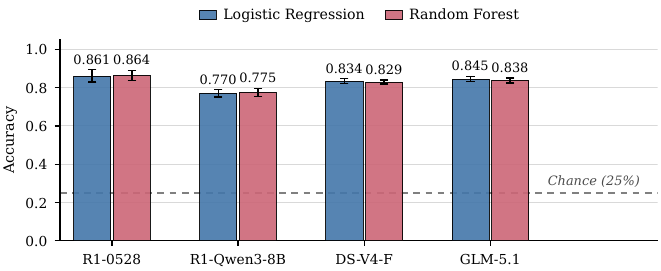}
\caption{Task Identification Accuracy Using General-Pattern Features. Error bars denote the standard deviation across cross-validation folds, and the dashed line indicates the 25\% random baseline.
Results for DS-V4-F and GLM-5.1 represent cross-model transfer using classifiers trained on R1-0528 traces.}
\label{fig:rq3-task-separability}
\end{figure}

\textbf{Results.}
Figure~\ref{fig:rq3-task-separability} presents task-identification accuracy using the 57 general-pattern features extracted from reasoning action traces.

\textit{1) Direct Results.}
Logistic regression achieves accuracies of $0.861\pm0.032$ for R1-0528 and $0.770\pm0.019$ for R1-Qwen3-8B, while random forest achieves $0.864\pm0.027$ and $0.775\pm0.022$, respectively. Both classifiers substantially exceed the 25\% random baseline for both models, showing that general-pattern features provide strong signals for distinguishing the four coding tasks. Because the classifiers use only features derived from reasoning action traces, these results show that task identity is reflected in the structural organization of LRM reasoning.

\textit{2) Transfer Results.}
The classifiers trained on R1-0528 also transfer to the two previously unseen models without retraining. On DS-V4-F, logistic regression and random forest achieve accuracies of $0.834\pm0.012$ and $0.829\pm0.011$, respectively. On GLM-5.1, they achieve $0.845\pm0.014$ and $0.838\pm0.013$. All four results substantially exceed the 25\% random baseline, showing that the task-discriminative signals captured by the general-pattern features transfer to target-model traces without model-specific adaptation.

\textit{3) Feature Importance Analysis.}
To further examine which aspects of the reasoning traces contribute to task identification, we analyze feature importance using mutual information. For the two primary R1 models, the most informative features primarily capture the allocation of reasoning effort across core actions, including the design ratio and the distributions of (PU), (SD), (IP), and (VV). These results suggest that action allocation provides a particularly informative signal for distinguishing coding tasks.

\begin{tcolorbox}[width=\linewidth, boxrule=0.8pt, left=2pt, right=2pt, top=2pt, bottom=2pt, colback=gray!10, colframe=black]
\textbf{Finding 3.1.}
General-pattern features support accurate coding-task identification, with action-allocation features providing the most informative signals. Classifiers trained on R1-0528 traces also transfer to DS-V4-F and GLM-5.1, achieving 0.829--0.845 accuracy without target-model retraining.
\end{tcolorbox}

\subsection{Pattern-Guided Prompting}
\label{sec:RQ3.2}

\textbf{Setup.}
RQ3.2 evaluates whether positive and negative reasoning patterns can be used as inference-time guidance for coding tasks. We compare four prompting conditions. \textit{Default} uses the original task instruction without additional reasoning guidance. \textit{Positive-only} augments the instruction with guidance derived solely from positive patterns mined from valid traces, whereas \textit{Negative-only} augments it with constraints derived solely from negative patterns mined from invalid traces. \textit{PN guidance} combines both positive-pattern guidance and negative-pattern constraints. All four conditions use the same task inputs and prompting settings, differing only in the reasoning guidance provided.

For the direct setting, we apply each primary R1 model's own pattern inventory to the corresponding model across the four coding tasks with objective correctness oracles. To evaluate cross-model transfer, we apply the pattern inventories mined from the R1 models to two previously unseen models from different model families, DeepSeek-V4-Flash (DS-V4-F)~\cite{deepseekv4} and GLM-5.1~\cite{glm51}. This design allows us to examine both within-model effectiveness and transferability of pattern-guided prompting.

We evaluate coding performance using pass@1~\cite{codex}. To reduce variance from stochastic decoding, each condition is evaluated in three independent runs, and we report the mean pass@1 across runs. We additionally report the mean number of generated output tokens per instance, measured in thousands, to characterize the effect of prompting on output length.

\begin{table}[t!]
\centering
\caption{Pattern-Guided Prompting Results and Cross-Model Transfer. \textit{Direct} applies each model's own pattern inventory, whereas \textit{Transferred} applies pattern inventories mined from the R1 models to the previously unseen models DS-V4-F and GLM-5.1. 
%P@1: Pass@1, measuring coding performance; 
Tok.: mean generated-output tokens per instance (thousands).
}
\label{tab:rq3-prompting}
\papertablestyle
\setlength{\tabcolsep}{6pt}
\renewcommand{\arraystretch}{1.0}
%\resizebox{\columnwidth}{!}{%
\begin{tabular}{ll *{8}{r}}
\toprule
& & \multicolumn{4}{c}{Direct} & \multicolumn{4}{c}{Transferred} \\
\cmidrule(lr){3-6}\cmidrule(lr){7-10}
Task & Method & \multicolumn{2}{c}{R1-0528} & \multicolumn{2}{c}{R1-Qwen3-8B} & \multicolumn{2}{c}{DS-V4-F} & \multicolumn{2}{c}{GLM-5.1} \\
\cmidrule(lr){3-4}\cmidrule(lr){5-6}\cmidrule(lr){7-8}\cmidrule(lr){9-10}
 &  & P@1$\uparrow$ & Tok.$\downarrow$ & P@1$\uparrow$ & Tok.$\downarrow$ & P@1$\uparrow$ & Tok.$\downarrow$ & P@1$\uparrow$ & Tok.$\downarrow$ \\
\midrule
\multirow{4}{*}{Generation}
 & Default & 72.46 & 14.9 & 56.21 & 17.9 & 80.78 & 13.0 & 90.29 & 16.5 \\
 & {Positive-only} & {73.14} & {15.2} & {55.86} & {18.3} & {84.36} & {13.5} & {90.06} & {13.2} \\
 & {Negative-only} & {72.08} & {14.6} & {56.39} & {17.6} & {79.84} & {12.5} & {90.74} & {13.0} \\
 & PN guidance & \textbf{74.49} & 15.6 & \textbf{56.88} & 17.3 & \textbf{87.80} & 12.2 & \textbf{91.17} & 12.9 \\
\midrule
\multirow{4}{*}{\makecell{Execution\\Reasoning}}
 & Default & 55.76 & 1.7 & 44.24 & 3.4 & 56.81 & 2.1 & 59.84 & 1.6 \\
 & {Positive-only} & {57.16} & {2.4} & {48.17} & {4.4} & {58.21} & {2.6} & {60.32} & {1.6} \\
 & {Negative-only} & {56.63} & {2.1} & {46.86} & {3.9} & {57.54} & {2.3} & {60.41} & {1.5} \\
 & PN guidance & \textbf{58.37} & 2.9 & \textbf{49.74} & 4.7 & \textbf{59.27} & 2.8 & \textbf{60.73} & 1.5 \\
\midrule
\multirow{4}{*}{Debugging}
 & Default & 88.80 & 4.5 & 61.20 & 7.7 & 91.40 & 2.4 & 93.00 & 2.8 \\
 & {Positive-only} & {91.20} & {5.8} & {61.54} & {7.8} & {90.92} & {2.5} & {93.02} & {2.6} \\
 & {Negative-only} & {90.47} & {5.1} & {61.08} & {7.6} & {92.18} & {2.2} & {93.11} & {2.6} \\
 & PN guidance & \textbf{94.60} & 6.6 & \textbf{61.72} & 7.9 & \textbf{92.80} & 2.3 & \textbf{93.20} & 2.5 \\
\midrule
\multirow{4}{*}{Translation}
 & Default & 55.08 & 5.7 & 39.09 & 4.2 & 55.05 & 2.1 & 56.97 & 2.5 \\
 & {Positive-only} & {56.81} & {6.5} & {42.51} & {4.4} & {54.73} & {2.2} & {58.03} & {2.4} \\
 & {Negative-only} & {57.44} & {6.2} & {43.27} & {4.3} & {56.73} & {1.8} & {58.52} & {2.4} \\
 & PN guidance & \textbf{58.78} & 7.3 & \textbf{44.14} & 4.5 & \textbf{57.44} & 1.9 & \textbf{58.75} & 2.3 \\
\bottomrule
\end{tabular}
%}
\end{table}

%\FloatBarrier

\textbf{Results.}
Table~\ref{tab:rq3-prompting} summarizes the effects of positive- and negative-pattern guidance on coding performance across the four tasks.

\textit{1) Direct Results.}
\textit{PN guidance} increases mean pass@1 on all four tasks for both R1-0528 and R1-Qwen3-8B. The largest gains are observed on debugging for R1-0528 (+5.80 percentage points) and on code execution reasoning and code translation for R1-Qwen3-8B (+5.50 and +5.05 points, respectively). The changes in performance are accompanied by varying changes in generated-output token usage. For example, on debugging, R1-0528 improves by 5.80 percentage points while mean generated-output tokens increase from 4.5K to 6.6K. On code generation, R1-Qwen3-8B improves by 0.67 points while generated-output tokens decrease slightly from 17.9K to 17.3K. These counts exclude the additional tokens introduced by the guidance prompt.

\textit{2) Ablation of Positive and Negative Guidance.}
We next compare the four prompting conditions. Across the sixteen task--model settings, neither \textit{Positive-only} nor \textit{Negative-only} consistently improves over \textit{Default} when used alone. \textit{Positive-only} reduces pass@1 for R1-Qwen3-8B on code generation (55.86 vs.\ 56.21) and for DS-V4-F on code translation (54.73 vs.\ 55.05). \textit{Negative-only} likewise reduces pass@1 for DS-V4-F on code generation (79.84 vs.\ 80.78) and for R1-Qwen3-8B on debugging (61.08 vs.\ 61.20). In contrast, \textit{PN guidance} achieves the highest pass@1 among the four conditions in every task--model setting and exceeds \textit{Default} in all sixteen settings.

\textit{3) Transfer Results.}
We further evaluate whether pattern inventories mined from the R1 models transfer to previously unseen models. On DS-V4-F, \textit{PN guidance} improves pass@1 on all four tasks, with the largest gain on code generation (+7.02 points). Generated-output token usage increases for code execution reasoning but decreases for the other three tasks. On GLM-5.1, pass@1 increases on all four tasks by 0.20--1.78 percentage points, while generated-output token usage decreases across all four tasks. For example, on code generation, pass@1 increases from 90.29 to 91.17 (+0.88 points), while mean generated-output tokens decrease from 16.5K to 12.9K.

\begin{tcolorbox}[width=\linewidth, boxrule=0.8pt, left=2pt, right=2pt, top=2pt, bottom=2pt, colback=gray!10, colframe=black]
\textbf{Finding 3.2.}
\textit{PN guidance} improves mean pass@1 across all evaluated task--model settings, including two previously unseen models. It also achieves higher pass@1 than the corresponding \textit{Positive-only} and \textit{Negative-only} variants in all sixteen settings.
\end{tcolorbox}

\subsection{Negative Pattern-Aware Early Stopping}
\label{sec:RQ3.3}

\textbf{Setup.} 
RQ3.3 evaluates whether negative-pattern signals can support online early stopping to reduce token usage while preserving coding accuracy. We compare two settings: \emph{Full Reasoning}, which allows the model to complete its reasoning trace before producing the final answer, and \emph{Pattern-Aware Stop}, which monitors the streaming reasoning trace and triggers answer finalization when a stopping condition is met.

Because \emph{Pattern-Aware Stop} requires streaming reasoning outputs, we evaluate it on R1-Qwen3-8B, DS-V4-F, and GLM-5.1, excluding R1-0528 because its interface does not provide streaming reasoning outputs. We directly apply the negative patterns mined from R1-Qwen3-8B to the same model and transfer them to DS-V4-F and GLM-5.1 without model-specific adaptation.

We run each setting three times and report mean pass@1 and mean API-reported token usage per instance. For stopped runs, token usage includes the finalizer request, which resubmits the original task with the generated reasoning prefix. The online monitor is rule-based and requires no additional model or annotation-model calls.

To quantify monitoring overhead, we replay 40 stored R1-Qwen3-8B traces comprising 9{,}019 chunks of 25 characters each. The monitor processes each chunk in 0.13\,ms at the median and 0.30\,ms at the 95th percentile, with cumulative CPU time of 19\,ms and 142\,ms per trace, respectively. This microbenchmark excludes model inference, network, and finalizer latency.

\begin{table}[t]
\centering
\caption{Negative Pattern-Aware Early Stopping Results across Models. Tok.: API-reported token usage (thousands); stopped runs include finalizer output.}
\label{tab:rq3-early-stopping}
\papertablestyle
\begin{threeparttable}
\setlength{\tabcolsep}{8pt}
\renewcommand{\arraystretch}{1.0}
\begin{tabular}{ll *{6}{r}}
\toprule
\multirow{2}{*}{Task} & \multirow{2}{*}{Model} & \multicolumn{2}{c}{Full Reasoning} & \multicolumn{4}{c}{Pattern-Aware Stop} \\
\cmidrule(lr){3-4}\cmidrule(lr){5-8}
 & & P@1$\uparrow$ & Tok.$\downarrow$ & P@1$\uparrow$ & Tok.$\downarrow$ & $\Delta$P@1 (pp) & $\Delta$Tok. (\%) \\
\midrule
\multirow{3}{*}{Generation}
 & R1-Qwen3-8B & 56.21 & 17.89K & 52.14 & 10.64K & \textcolor{red!70!black}{$-4.07$} & \textcolor{green!50!black}{$-40.56\%$} \\
 & DS-V4-F & 80.78 & 13.00K & 73.36 & 8.43K & \textcolor{red!70!black}{$-7.42$} & \textcolor{green!50!black}{$-35.13\%$} \\
 & GLM-5.1 & 90.29 & 16.54K & 88.06 & 7.85K & \textcolor{red!70!black}{$-2.23$} & \textcolor{green!50!black}{$-52.52\%$} \\
\midrule
\multirow{3}{*}{\makecell[l]{Execution\\Reasoning}}
 & R1-Qwen3-8B & 44.24 & 3.43K & 40.58 & 2.10K & \textcolor{red!70!black}{$-3.66$} & \textcolor{green!50!black}{$-38.75\%$} \\
 & DS-V4-F & 56.81 & 2.13K & 51.57 & 0.85K & \textcolor{red!70!black}{$-5.24$} & \textcolor{green!50!black}{$-59.78\%$} \\
 & GLM-5.1 & 59.84 & 1.64K & 58.63 & 0.52K & \textcolor{red!70!black}{$-1.21$} & \textcolor{green!50!black}{$-68.53\%$} \\
\midrule
\multirow{3}{*}{Debugging}
 & R1-Qwen3-8B & 61.20 & 7.65K & 54.40 & 4.18K & \textcolor{red!70!black}{$-6.80$} & \textcolor{green!50!black}{$-45.32\%$} \\
 & DS-V4-F & 91.40 & 2.42K & 86.60 & 1.69K & \textcolor{red!70!black}{$-4.80$} & \textcolor{green!50!black}{$-30.16\%$} \\
 & GLM-5.1 & 93.00 & 2.77K & 91.20 & 1.74K & \textcolor{red!70!black}{$-1.80$} & \textcolor{green!50!black}{$-37.16\%$} \\
\midrule
\multirow{3}{*}{Translation}
 & R1-Qwen3-8B & 39.09 & 4.18K & 37.64 & 2.33K & \textcolor{red!70!black}{$-1.45$} & \textcolor{green!50!black}{$-44.22\%$} \\
 & DS-V4-F & 55.05 & 2.14K & 50.79 & 1.23K & \textcolor{red!70!black}{$-4.26$} & \textcolor{green!50!black}{$-42.65\%$} \\
 & GLM-5.1 & 56.97 & 2.53K & 55.48 & 1.51K & \textcolor{red!70!black}{$-1.49$} & \textcolor{green!50!black}{$-40.56\%$} \\
\bottomrule
\end{tabular}
%\vspace{2pt}
%\begin{tablenotes}[flushleft]
%\item[] \textcolor{blue}{Tok.: API-reported token usage (thousands); stopped runs include finalizer output.}
%\end{tablenotes}
\end{threeparttable}
\end{table}

%\FloatBarrier

\textbf{Results.}
Table~\ref{tab:rq3-early-stopping} summarizes the effects of negative pattern-aware early stopping across the four coding tasks.

\textit{1) Direct Results.}
For R1-Qwen3-8B, API-reported token usage decreases by 38.75\%--45.32\%, accompanied by a 1.45--6.80 percentage-point decrease in pass@1. Code translation shows the smallest accuracy decrease (1.45 points) with a 44.22\% reduction in token usage, whereas code generation and debugging show larger accuracy decreases.

\textit{2) Transfer Results.}
Applying the negative patterns mined from R1-Qwen3-8B to the two unseen models also reduces API-reported token usage. The reductions range from 30.16\% to 59.78\% for DS-V4-F and from 37.16\% to 68.53\% for GLM-5.1, with corresponding pass@1 decreases of 4.26--7.42 and 1.21--2.23 percentage points, respectively. 
Thus, the same negative-pattern signals reduce token usage across all three models, although the associated accuracy changes vary by task and model.

\begin{tcolorbox}[width=\linewidth, boxrule=0.8pt, left=2pt, right=2pt, top=2pt, bottom=2pt, colback=gray!10, colframe=black]
\textbf{Finding 3.3.}
Negative pattern-aware early stopping reduces API-reported token usage by 30--69\% across the three evaluated models, with a corresponding 1.21--7.42 percentage-point decrease in pass@1 across task--model settings.
\end{tcolorbox}

\section{Discussion}

\subsection{Understanding Reasoning Structures}

The patterns suggest that reasoning structures arise from how a task constrains uncertainty and when the model can commit to an intermediate result. All four tasks begin with problem understanding because the model must first establish what is being transformed, executed, repaired, or produced. They then diverge because their uncertainty is resolved differently: generation and translation require constructing a representation before it can be checked, whereas execution and debugging require maintaining an externally constrained state and repeatedly testing that state. The resulting task-specific structures therefore reflect different control demands, rather than superficial differences in trace length or vocabulary.

The validity contrasts reveal what happens when this control process loses its anchor. Verification is not merely another action; it provides a checkpoint that can confirm or reject the current interpretation and thereby allow the trace to move forward. Design, reflection, or knowledge retrieval can be useful when they reduce a concrete uncertainty, but repeated transitions among them without renewed verification reopen decisions that have not been resolved. This explains why longer reasoning is not a reliable sign of better reasoning: additional steps help only when they convert uncertainty into a checked commitment. Positive and negative patterns consequently distinguish two forms of iteration, one that closes a loop through evidence and one that perpetuates exploration without closure.

This interpretation also explains the cross-model results. Models can realize the same task-level strategy with different amounts of repetition and dispersion because the underlying task constraints remain stable while the efficiency of the control process varies. The structural representation separates these two factors: general patterns describe the task-conditioned workflow, while local transitions expose whether the workflow is being stabilized by verification or repeatedly reopened. Reasoning structure is therefore a process-level signal of how a model manages uncertainty and commits to intermediate states, which is why it can support task identification and inference-time intervention without treating every additional reasoning token as useful progress.

\subsection{Future Research Directions}

\textbf{Evaluation of Reasoning Models.}
Reasoning traces provide behavioral signals not captured by outcome-based metrics such as pass@1 or trace length. Future evaluation frameworks could incorporate action allocation and transition structure to provide a finer-grained characterization of reasoning behavior.

\textbf{Structure-Aware Inference Control.}
The discovered patterns can serve as structural signals for inference-time monitoring and intervention. Beyond early stopping, such signals could support adaptive reasoning budgets, targeted verification, or re-planning when recurring drift patterns are detected.

\textbf{Learning with Reasoning Structure.}
Beyond inference-time guidance, positive and negative patterns could provide process-level supervision for training. Training objectives could reward desirable action transitions and discourage recurring negative patterns, moving from externally guided reasoning toward models that learn structured reasoning behavior.

\section{Threats to Validity}

\textbf{Construct Validity.}
We abstract free-form reasoning traces into discrete reasoning actions to enable systematic comparison, which may omit fine-grained semantic distinctions. We mitigate this risk by grounding the taxonomy in prior work, refining it through expert inspection, and validating automatic annotations against human re-labeling (Cohen's $\kappa=0.816$). Nevertheless, alternative taxonomies, segmentation granularities, or collapsing rules may yield different patterns.

\textbf{Internal Validity.}
Trace-validity labels are derived from an ensemble of LLM judges, with human adjudication for low-confidence cases. Confidence-weighted voting reduces judgment variance, but subtle reasoning errors may remain undetected. Task-level functional correctness is evaluated independently using objective oracles and reported as pass@1. The prompting and early-stopping results therefore provide intervention evidence but do not establish causal effects attributable to individual mined patterns.

For early stopping, the online rule-based mapper is distinct from the offline LLM-based annotators used for pattern discovery. Their agreement on partial reasoning prefixes has not been independently evaluated, so mapping errors may affect motif detection and stopping decisions. The stopping policy also includes a repeated-text trigger that operates independently of action motifs.

\textbf{External Validity.}
We study four coding tasks and four model configurations, which may not capture the diversity of software engineering tasks or model architectures. R1-0528 and its Qwen3-8B distillation share a model lineage, while DS-V4-F is from the same model family; GLM-5.1 provides one cross-family evaluation. Moreover, GLM-5.1 also participates in the validity-judge ensemble for the primary traces, so its intervention results do not constitute a fully independent replication of the discovery-to-intervention pipeline. Broader evaluation across tasks, programming languages, and model families is needed to assess the generality of the findings.

\section{Related Work}

Recent LRMs incorporate extended reasoning traces as a core component of inference~\cite{deepseekr1}. In code-related settings, prior work shows that reasoning-before-coding is not uniformly beneficial and depends on how reasoning is structured relative to code generation~\cite{revisitcotcodegen}. Other studies explore leveraging external or structured signals to improve reasoning, such as mining discussion data into training chains~\cite{svrc2026} or validating intermediate steps via search-based refinement~\cite{mctsrefine2025}. These efforts demonstrate the utility of explicit reasoning for code generation, but primarily focus on improving outputs or training signals rather than characterizing recurring reasoning structures across tasks.

Prior work has examined reasoning quality and failure modes. Studies find that reasoning quality does not always correspond to final correctness: flawed reasoning can still lead to correct outputs, while correct outputs can also arise from flawed reasoning~\cite{cotcodequality}. Similarly, surface-level properties such as reasoning length or step count provide weak evidence of correctness~\cite{reasoningsteps2026}. Recent studies further categorize reasoning behaviors and failure patterns, including hallucination, overthinking, and erroneous backtracking~\cite{thinkingpatterns,reasoninghallucination2025,demystifyingerrors2025}. While these studies move beyond outcome-based evaluation, they often focus on individual tasks or rely on coarse-grained or manually defined categories. How reasoning structures recur across coding tasks, and how such structures distinguish valid from invalid reasoning traces, therefore remains less explored.

A separate line of work uses reasoning traces for inference-time control. SEER adapts reasoning strategies during generation~\cite{seer2025}, while studies of overthinking motivate methods for removing redundant reasoning steps~\cite{overthinking2025,asap2025,adaptivecot2025}. Other early-stopping approaches use intermediate signals, such as answer convergence or confidence, to terminate generation earlier~\cite{answerconvergence2025,deer2026}. These approaches primarily target decoding efficiency or reasoning control, but do not derive the control signals from recurring reasoning structures identified through empirical analysis of traces.

In contrast, we treat reasoning traces as behavioral data and abstract them into reasoning action traces to study recurring reasoning structures across coding tasks. This allows us to examine reasoning patterns at both the task and action levels, with particular attention to their recurrence across tasks and their association with trace validity.

\section{Conclusion}
In this paper, we present an empirical study of reasoning patterns in AI coding. By abstracting reasoning traces into action sequences, we identify general patterns that capture task-specific reasoning strategies and positive and negative patterns that characterize valid and invalid reasoning behaviors at the action level across four coding tasks. We further demonstrate the utility of these patterns for inference-time analysis and control through task identification, pattern-guided prompting, and negative pattern-aware early stopping. These results show that reasoning patterns can provide a structured basis for analyzing LRM reasoning and controlling inference-time behavior in AI coding.

\section*{Data Availability}

All code and data used in this study are publicly available at: \url{https://anonymous.4open.science/r/FSE_2027}. The repository includes: (i) the collected reasoning traces and annotated datasets across the four tasks (code generation, code execution reasoning, debugging, and translation); (ii) the code for  pattern discovery, pattern-guided prompting and pattern-aware early stopping (RQ1--RQ3); and (iii) all experimental results and evaluation scripts.

% \clearpage
\bibliographystyle{ACM-Reference-Format}
\bibliography{ref}

\end{document}